\documentclass[twocolumn,english,floatfix]{revtex4}
\usepackage[T1]{fontenc}
\usepackage[latin9]{inputenc}
\usepackage{amsmath}
\usepackage{amssymb}
\usepackage{graphicx}
\usepackage{esint}

\makeatletter
\@ifundefined{textcolor}{}
{%
 \definecolor{BLACK}{gray}{0}
 \definecolor{WHITE}{gray}{1}
 \definecolor{RED}{rgb}{1,0,0}
 \definecolor{GREEN}{rgb}{0,1,0}
 \definecolor{BLUE}{rgb}{0,0,1}
 \definecolor{CYAN}{cmyk}{1,0,0,0}
 \definecolor{MAGENTA}{cmyk}{0,1,0,0}
 \definecolor{YELLOW}{cmyk}{0,0,1,0}
}

\usepackage{babel}

\usepackage{babel}

\makeatother

\usepackage{babel}

\makeatother

\usepackage{babel}
\begin{document}

\title{Impact of local-moment fluctuations on the magnetic degeneracy of
iron arsenide superconductors}

\author{Xiaoyu Wang and Rafael M. Fernandes}

\affiliation{School of Physics and Astronomy, University of Minnesota, Minneapolis,
MN 55455, USA}

\date{\today }
\begin{abstract}
We investigate the fate of the orthorhombic stripe-type magnetic state
(ordering vectors $\left(\pi,0\right)$/$\left(0,\pi\right)$), observed
in most iron-pnictide superconductors, in the presence of localized
magnetic moments that tend to form a Néel state (ordering vector $\left(\pi,\pi\right)$).
We show that before long-range Néel order sets in, the coupling between
the conduction electrons and the fluctuations of the local moments
favors an unusual magnetic state consisting of a coherent superposition
of the $\left(\pi,0\right)$ and $\left(0,\pi\right)$ orders that
preserves tetragonal symmetry. The magnetization of this state is
non-uniform and induces a simultaneous checkerboard charge order.
We discuss signatures of this magnetic configuration on the electronic
spectrum and its impact on the superconducting state, showing that
its phase space for coexistence with the $s^{+-}$ state is smaller
than the stripe-type state. Our results shed light on recent experimental
observations on $\mathrm{Ba(Fe_{1-x}Mn_{x})_{2}As_{2}}$ compounds,
where the Néel-type local Mn moments interact with the Fe conduction
electrons. 
\end{abstract}


\maketitle

\section{Introduction}

The proximity between magnetic order and unconventional superconductivity
in several materials has been a key motivation to investigate pairing
mediated by spin fluctuations \cite{Scalapino_RMP}. Interestingly,
the parent compounds of the two families of high-temperature superconductors,
cuprates and iron pnictides, display rather different magnetic ground
states. While in the former a Mott insulating Néel-type magnetic configuration
(ordering vector $\mathbf{Q}_{N}=\left(\pi,\pi\right)$) is observed,
in the latter one finds a metallic stripe-type state (ordering vectors
$\mathbf{Q}_{1}=\left(\pi,0\right)$ or $\mathbf{Q}_{2}=\left(0,\pi\right)$
in the Fe-square lattice) that breaks the tetragonal symmetry of the
system down to orthorhombic. These differences in the magnetic spectra
are manifested in the distinct pairing states promoted by the spin
fluctuations -- d-wave for the cuprates and $s^{+-}$ for the iron
pnictides \cite{reviews_pairing}.

To better understand the similarities and differences between these
two classes of materials, it is desirable to study a system that interpolates
between these two magnetic ground states \cite{LaMnPO_Kotliar}. Experimentally,
a promising material is the $\mathrm{Ba(Fe_{1-x}Mn_{x})_{2}As_{2}}$
compound: for $x=0$ it undergoes a nearly simultaneous magnetic-structural
transition to a metallic stripe-type state at $T_{\mathrm{mag}}\approx137$K
\cite{Kim11,Birgeneau11} with a saturated magnetic moment of about
$0.9\mu_{B}$ \cite{reviews}. Optical conductivity \cite{Uchida10}
and ARPES \cite{Liu10} measurements indicate that the conduction
Fe electrons are directly involved in the formation of the magnetic
state, in agreement with first-principle calculations \cite{Andersen11}.
For $x=1$ the system undergoes a magnetic transition at much higher
temperatures, $T_{\mathrm{mag}}\approx625$K, forming an insulating
Néel state with a large saturated magnetic moment of $3.9\mu_{B}$
\cite{Mn_synthesis,Mn_electronic_structure,Mn_K_doped_ARPES}. Whether
this state is a Mott insulator remains to be seen \cite{Mn_correlations_Imada},
but both theory and experiment suggest that correlations are stronger
than in the $x=0$ compound \cite{Mn_correlations_Schmalian}, and
that a local Mn moment picture describes well the ordered state \cite{Mn_Heisenberg_model,Mn_NMR,Mn_localized_photoemission,Mn_Mossbauer,Mn_doped_1111}.
Although no superconductivity has been observed in these compounds,
short-range Néel fluctuations, presumably arising from the Mn moments,
are observed via neutron scattering even for small doping levels $x\approx0.07$
\cite{Mn_neutron}. Remarkably, x-ray and neutron diffraction measurements
report an unusual intermediate magnetic state for $x\gtrsim0.1$,
which does not break the tetragonal symmetry of the system despite
the presence of magnetic Bragg peaks at $\mathbf{Q}_{1}=\left(\pi,0\right)$
or $\mathbf{Q}_{2}=\left(0,\pi\right)$ \cite{Kim10}.

\begin{figure}
\begin{centering}
\includegraphics[width=0.8\columnwidth]{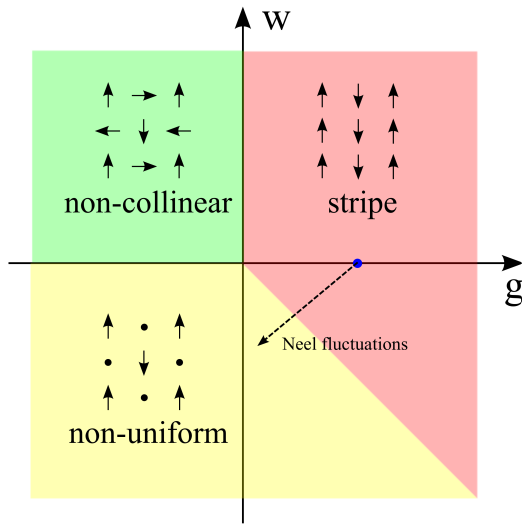} 
\par\end{centering}

\caption{Phase diagram of the Ginzburg-Landau model (\ref{F_general}), displaying
the orthorhombic stripe-type state for $g>\mathrm{max}\left\{ 0,-w\right\} $,
as well as the tetragonal non-uniform and non-collinear states for
$g<\mathrm{max}\left\{ 0,-w\right\} $ (see also Refs. \cite{Lorenzana08,Eremin10}).
In the absence of Néel fluctuations, the ground state is the stripe
one (blue dot). As Néel fluctuations increase, a transition from the
stripe to the non-uniform state takes place (dashed arrow). \label{fig_phase_diagram}}
\end{figure}

Theoretically, the transition from a stripe phase to a Néel state
may seem at first sight straightforward. In a square-lattice local-moment
model with nearest-neighbor and next-nearest-neighbor antiferromagnetic
exchanges $J_{1}$ and $J_{2}$, respectively, there is a classical
transition from a stripe to a Néel state once $J_{1}>2J_{2}$ \cite{Chandra90}.
However, the fact that the stripe state in the pnictides is metallic,
with conduction electrons forming the magnetic moments, opens novel
possibilities. This is because the itinerant magnetic state driven
by the nesting properties of the Fermi surface is highly degenerate
\cite{Lorenzana08,Eremin10,Tesanovic11,Brydon11,Fernandes12}: besides
the stripe phase, other configurations that do not break tetragonal
symmetry, with non-collinear or non-uniform magnetization, may minimize
the magnetic free energy (see Fig. \ref{fig_phase_diagram}). The
interaction with local moments affects this intricate free energy
landscape, and can potentially give rise to unusual magnetic ground
states.

In this paper, motivated by the physics of these $\mathrm{Ba(Fe_{1-x}Mn_{x})_{2}As_{2}}$
compounds, we show that short-range Néel-type fluctuations favor a
different magnetic state that does not break tetragonal symmetry but
that still displays magnetic Bragg peaks at $\mathbf{Q}_{1}=\left(\pi,0\right)$
and $\mathbf{Q}_{2}=\left(0,\pi\right)$, in qualitative agreement
with the observations in the $x\approx0.1$ $\mathrm{Ba(Fe_{1-x}Mn_{x})_{2}As_{2}}$
compounds. Its magnetic configuration is non-uniform, inducing a secondary
charge density-wave with ordering vector $\mathbf{Q}_{N}$, which
can be detected experimentally. We also determine the changes in the
electronic spectrum -- which can be probed by ARPES and STM -- promoted
by this magnetic tetragonal state. The main difference from the reconstructed
Fermi surface of the stripe state is the absence of a central unhybridized
hole pocket, replaced by additional reconstructed pockets at high-symmetry
directions. Finally, we show that the non-uniform state tends to phase-separate
from the $s^{+-}$ superconducting state, which helps to explain the
absence of coexisting superconductivity in the $\mathrm{Ba(Fe_{1-x}Mn_{x})_{2}As_{2}}$
compounds, in contrast to their $\mathrm{Ba(Fe_{1-x}Co_{x})_{2}As_{2}}$
counterparts.

The paper is organized as follows: in Section II we develop a general
Ginzburg-Landau model that captures the three different possible magnetic
ground states of the iron pnictides. In Section III we introduce a
microscopic model where the itinerant electrons couple to local Néel
moments, showing that Néel fluctuations favor the non-uniform tetragonal
magnetic state. In Section IV we discuss the reconstructed electronic
spectrum due to this peculiar order, and in Section V, its impact
on superconductivity. Section VI is devoted to the concluding remarks.

\section{Phenomenological model: degeneracy of the magnetic ground state}

The enlarged degeneracy of the itinerant magnetic ground state of
the iron pnictides can be captured by a phenomenological Ginzburg-Landau
model \cite{Lorenzana08,Brydon11,Fernandes12}. In the tetragonal
phase, neutron scattering experiments find magnetic fluctuations of
equal amplitude peaked at the two ordering vectors $\mathbf{Q}_{1}=\left(\pi,0\right)$
and $\mathbf{Q}_{2}=\left(0,\pi\right)$ \cite{Mn_neutron}. Therefore,
we introduce two $O(3)$ magnetic order parameters, $\mathbf{M}_{1}$
and $\mathbf{M}_{2}$, associated respectively with $\mathbf{Q}_{1}$
and $\mathbf{Q}_{2}$. As a result, the spin at position $\mathbf{r}$
is in general a superposition of the two order parameters, i.e. $\mathbf{S}\left(\mathbf{r}\right)=\mathbf{M}_{1}\mathrm{e}^{i\mathbf{Q}_{1}\cdot\mathbf{r}}+\mathbf{M}_{2}\mathrm{e}^{i\mathbf{Q}_{2}\cdot\mathbf{r}}$.
The most general free energy expansion that respects tetragonal and
$O(3)$ symmetries is:

\begin{eqnarray}
F & = & \frac{a}{2}\left(M_{1}^{2}+M_{2}^{2}\right)+\frac{u}{4}\left(M_{1}^{2}+M_{2}^{2}\right)^{2}\nonumber \\
 &  & -\frac{g}{4}\left(M_{1}^{2}-M_{2}^{2}\right)^{2}+w\left(\mathbf{M}_{1}\cdot\mathbf{M}_{2}\right)^{2}\label{F_general}
\end{eqnarray}

The first two terms depend only on the combination $M_{1}^{2}+M_{2}^{2}$,
effectively enlarging the symmetry of the system to $O(6)$, and resulting
in a huge degeneracy of the magnetic ground state \cite{Eremin10,Tesanovic11}.
The last two terms are responsible for lifting this degeneracy, selecting
both the relative amplitudes (either $M_{1}^{2}/M_{2}^{2}=0$ or $M_{1}^{2}/M_{2}^{2}=1$)
and the relative orientations of the two order parameters (either
$\mathbf{M}_{1}\parallel\mathbf{M}_{2}$ or $\mathbf{M}_{1}\perp\mathbf{M}_{2}$).
Fig. 1 displays the phase diagram and the resulting ground states
as function of $g$ and $w$. For $g>\mathrm{max}\left\{ 0,-w\right\} $,
we find a stripe-type state, characterized by $M_{1}\neq0$ and $M_{2}=0$
(or vice-versa), which breaks the tetragonal symmetry of the system.
This is the state most commonly observed in the iron pnictides and
has a residual $Z_{2}$ (Ising) symmetry, related to choosing either
$M_{1}\neq0$ or $M_{2}\neq0$, which can be broken even before the
magnetic transition takes place \cite{Fernandes12}.

For $g<\mathrm{max}\left\{ 0,-w\right\} $, minimization of the free
energy leads to a tetragonal magnetic state characterized by simultaneously
non-vanishing $M_{1}^{2}=M_{2}^{2}$. Two different configurations
are possible: for $w>0$, we obtain the non-collinear state $\mathbf{M}_{1}\perp\mathbf{M}_{2}$,
where nearest-neighbor spins of amplitude $\left\langle S_{i}\right\rangle =\sqrt{2}M$
are orthogonal to each other (see Fig. \ref{fig_phase_diagram} and
also Refs. \cite{Lorenzana08,Eremin10}). For $w<0$, the ground state
is given by $\mathbf{M}_{1}\parallel\mathbf{M}_{2}$, corresponding
to a non-uniform collinear state (see Fig. \ref{fig_phase_diagram}
and Refs. \cite{Lorenzana08,Eremin10}). In this configuration, odd
sites of the original square lattice form a non-magnetic sublattice,
with local spin $\left\langle S_{i_{\mathrm{odd}}}\right\rangle =0$,
whereas even sites form a Néel sublattice, with $\left\langle S_{i_{\mathrm{even}}}\right\rangle =2M$.
This non-uniform state induces a charge density-wave (CDW) with modulation
$\mathbf{Q}_{1}+\mathbf{Q}_{2}=\mathbf{Q}_{N}$, where the odd (non-magnetic)
sites have different local charge than the even (magnetic) sites.
This can be obtained from Eq. (\ref{F_general}) by including the
charge degrees of freedom \cite{Balatsky10}:

\begin{equation}
\tilde{F}=F-\zeta\,\rho_{\mathbf{Q}_{N}}\left(\mathbf{M}_{1}\cdot\mathbf{M}_{2}\right)+\frac{1}{2}\,\chi_{\mathrm{CDW}}^{-1}\rho_{\mathbf{Q}_{N}}^{2}\label{F_tilde}
\end{equation}

Here, $\rho_{\mathbf{Q}_{N}}$ is the Fourier component of the charge
density $\rho\left(\mathbf{r}\right)$ at momentum $\mathbf{Q}_{N}=\left(\pi,\pi\right)$,
i.e. it is related to a checkerboard charge-density wave. Minimization
with respect to the CDW order parameter gives $\rho_{\mathbf{Q}_{N}}=\chi_{\mathrm{CDW}}\zeta\left(\mathbf{M}_{1}\cdot\mathbf{M}_{2}\right)$,
implying that its amplitude in the magnetically ordered state depends
on both the coupling constant $\zeta$ and the bare CDW susceptibility
$\chi_{\mathrm{CDW}}$.

The vast majority of iron pnictides display a stripe-type ground state,
$g>\mathrm{max}\left\{ 0,-w\right\} $. The recent observation in
the $\mathrm{Ba(Fe_{1-x}Mn_{x})_{2}As_{2}}$ compounds of a magnetic
state with peaks at $\mathbf{Q}_{1}=\left(\pi,0\right)$ and $\mathbf{Q}_{2}=\left(0,\pi\right)$
but no orthorhombic distortion \cite{Kim10} indicates that upon sufficient
Mn doping, the Ginzburg-Landau coefficients change and bring the system
to the regime of tetragonal magnetic states (either the non-uniform
or the non-collinear state). Our goal now is to develop a microscopic
model to evaluate these coefficients and unveil the mechanism behind
these changes.

\section{Microscopic model: impact of Néel fluctuations}

The typical Fermi surface of the iron pnictides is shown in Fig. \ref{fig_reconstructed_FS}(a),
obtained from the tight-binding model of Ref. \cite{Graser09}. To
make our analysis more transparent, we follow Refs. \cite{Eremin10,Fernandes12}
and consider an effective model with a (possibly degenerate) circular
hole pocket $h$ at the center of the Brillouin zone and two elliptical
electron pockets $e_{1,2}$ centered at $\mathbf{Q}_{1}=\left(\pi,0\right)$
and $\mathbf{Q}_{2}=\left(0,\pi\right)$. The band dispersions are
respectively, 
\[
\begin{split}\varepsilon_{h,\mathbf{k}} & =\epsilon_{0}-\frac{k^{2}}{2m}-\mu\\
\varepsilon_{e_{1},\mathbf{k}+\mathbf{Q}_{1}} & =-\epsilon_{0}+\frac{k_{x}^{2}}{2m_{x}}+\frac{k_{y}^{2}}{2m_{y}}-\mu\\
\varepsilon_{e_{2},\mathbf{k}+\mathbf{Q}_{2}} & =-\epsilon_{0}+\frac{k_{x}^{2}}{2m_{y}}+\frac{k_{y}^{2}}{2m_{x}}-\mu
\end{split}
\]

Close to particle-hole symmetry (perfect nesting), we can rewrite
the band dispersions in a more convenient form:

\begin{equation}
\begin{split}\varepsilon_{h,\mathbf{k}} & =-\varepsilon_{\mathbf{k}}\\
\varepsilon_{e_{1},\mathbf{k}+\mathbf{Q}_{1}} & =\varepsilon_{\mathbf{k}}-(\delta_{\mu}+\delta_{m}\cos2\theta)\\
\varepsilon_{e_{2},\mathbf{k}+\mathbf{Q}_{2}} & =\varepsilon_{\mathbf{k}}-(\delta_{\mu}-\delta_{m}\cos2\theta)
\end{split}
\end{equation}
where $\theta$ is the angle around the Fermi surface. The parameter
$\delta_{\mu}$ is related to the occupation number (doping) and $\delta_{m}$,
to the ellipticity of the electron pockets:

\begin{eqnarray}
\delta_{\mu} & = & 2\mu+\epsilon_{F}\left[1-\frac{m}{2}\left(\frac{m_{x}+m_{y}}{m_{x}m_{y}}\right)\right]\nonumber \\
\delta_{m} & = & \frac{\epsilon_{F}m}{2}\left(\frac{m_{x}-m_{y}}{m_{x}m_{y}}\right)\label{parameters}
\end{eqnarray}
where $\epsilon_{F}$ is the Fermi energy. Thus, the non-interacting
Hamiltonian is given by $H_{0}=\sum_{\mathbf{k},a}\varepsilon_{\mathbf{k},a}c_{a,\mathbf{k}\alpha}^{\dagger}c_{a,\mathbf{k}\alpha}^{\phantom{\dagger}}$,
with band index $a$ and spin index $\alpha$. Projecting the interacting
Hamiltonian in the SDW channel \cite{Maiti10} -- which is the leading
instability of the system -- yields the term $H_{I}=U_{\mathrm{SDW}}\sum_{\mathbf{q},i}\mathbf{s}_{\mathbf{q}}^{(i)}\cdot\mathbf{s}_{-\mathbf{q}}^{(i)}$
where $\mathbf{s}_{\mathbf{q}}^{(i)}=\sum_{\mathbf{k}}c_{h,\mathbf{k+q}\alpha}^{\dagger}\boldsymbol{\sigma}_{\alpha\beta}c_{e_{i},\mathbf{k}+\mathbf{Q}_{i}\beta}^{\phantom{\dagger}}$
are the two staggered spin operators whose mean values are proportional
to the two order parameters $\mathbf{M}_{i}$.

The free energy (\ref{F_general}) can now be \emph{derived }from
the total Hamiltonian $H_{0}+H_{I}$ by performing Hubbard-Stratonovich
transformations and integrating out the electronic degrees of freedom
\cite{Fernandes12}. We consider $\mathbf{M}_{i}$ to be real and
homogeneous. The Ginzburg-Landau coefficients, as obtained in Ref.
\cite{Fernandes12}, are given by $w=0$ and: 
\begin{align}
u & =\int_{k}G_{h,k}^{2}\left(G_{e_{1},k}+G_{e_{2},k}\right)^{2}\approx\frac{7\zeta(3)\rho_{F}}{2\pi^{2}T^{2}}\\
g & =-\int_{k}G_{h,k}^{2}\left(G_{e_{1},k}-G_{e_{2},k}\right)^{2}\approx\frac{31\zeta(5)\rho_{F}}{32\pi^{4}T^{2}}\left(\frac{\delta_{m}}{T}\right)^{2}
\end{align}
 where $G_{a,k}^{-1}=i\omega_{n}-\varepsilon_{a,\mathbf{k}}$ are
the non-interacting single-particle Green's functions, and $\rho_{F}$
is the density of states at the Fermi level. In the limit of perfect
nesting (i.e. $\delta_{\mu}=\delta_{m}=0$) one obtains $g=w=0$,
implying that the system has an enlarged $O(6)$ symmetry and a huge
ground-state degeneracy. Expanding near perfect nesting yields $g\propto\delta_{m}^{2}>0$
and $w=0$, placing the system in the regime of a stripe-type magnetic
state (blue dot in Fig. \ref{fig_phase_diagram}). Similar free energy
calculations considering other types of band dispersions also find
that the stripe state is favored for a wide range of parameters, consistent
with the observations that most iron pnictides display this magnetic
ordered state \cite{Lorenzana08,Brydon11,Bascones_phase_diagram,Dagotto_phase_diagram}.

To make contact with the $\mathrm{Ba(Fe_{1-x}Mn_{x})_{2}As_{2}}$
compounds, we include the coupling between the conduction electrons
and Néel-type fluctuations. As shown by first-principle and model
calculations, Néel fluctuations are always present in the iron pnictides
due to the existence of two matching electron pockets separated by
$\mathbf{Q}_{N}=\left(\pi,\pi\right)$ \cite{Johannes09,Bascones_phase_diagram,Fernandes13}.
The presence of Mn dopants enhances these fluctuations, because the
magnetic Mn dopants promote Néel order -- indeed, the ``fully doped''
$\mathrm{BaMn_{2}As_{2}}$ compound displays a transition to a Néel
magnetic state at rather high temperatures \cite{Mn_synthesis}. The
coupling between the local Mn moments and the Fe conduction electrons
is attested by local probes such as ESR (electron spin resonance)
\cite{ESR} and NMR (nuclear magnetic resonance) \cite{Mn_NMR}. This
unique behavior of the Mn dopants should be contrasted with other
types of chemical substitution in the Fe site, $\mathrm{Ba(Fe_{1-x}}M\mathrm{_{x})_{2}As_{2}}$,
such as $M=$ Co, Ni, Cu. For instance, Co and Ni are non-magnetic,
as shown by ESR measurements \cite{ESR}. Cu, although magnetic, does
not seem to favor a Néel state, since the ``fully doped'' $\mathrm{BaCu_{2}As_{2}}$
compound remains paramagnetic \cite{BaCu2As2_1,BaCu2As2_2}.

Experimental evidence for Néel fluctuations in $\mathrm{Ba(Fe_{1-x}Mn_{x})_{2}As_{2}}$
is given by neutron diffraction experiments, which observe an inelastic
magnetic peak at $\mathbf{Q}_{N}=\left(\pi,\pi\right)$ already for
very small Mn-doping levels $x$ \cite{Mn_neutron}, where no long-range
Néel order is observed. In this dilute limit, Mn dopants are also
expected to promote impurity scattering. One of its main effects is
to suppress the magnetic transition temperature $T_{\mathrm{mag}}$,
as discussed in Refs. \cite{disorder1,disorder2}. Within our model,
$T_{\mathrm{mag}}$ appears in the quadratic term of the general Ginzburg-Landau
expansion in Eq. (\ref{F_general}), and therefore is not responsible
for the selection of the ground state -- which is determined solely
by the quartic coefficients. Thus, hereafter we focus only on the
role played by Néel fluctuations. Denoting by $\mathbf{N}$ the collective
field associated with these Néel fluctuations, and by $\chi_{N}\left(\mathbf{q}\right)$
the corresponding momentum-dependent susceptibility, we therefore
consider the coupling between the Néel fluctuations and the itinerant
electrons according to \cite{Fernandes13}:

\begin{equation}
H_{N}=\sum_{\mathbf{k}}\mathbf{N}\cdot\left(c_{e_{1},\mathbf{k+Q}_{1}\alpha}^{\dagger}\boldsymbol{\sigma}_{\alpha\beta}c_{e_{2},\mathbf{k}+\mathbf{Q}_{2}\beta}^{\phantom{\dagger}}\right)\label{H_loc}
\end{equation}
where, for simplicity, the coupling constant was incorporated to $\mathbf{N}$.
To determine how the magnetic ground state is affected by short-range
Néel fluctuations, we rederive the coefficients of the free energy
(\ref{F_general}) from the Hamiltonian $H_{0}+H_{I}+H_{N}$, expanding
to the leading quadratic order in $\mathbf{N}$: 
\begin{equation}
\begin{split}\delta F= & \frac{\alpha}{2}N^{2}\left(M_{1}^{2}+M_{2}^{2}\right)-4\lambda_{12}\left[\left(\mathbf{M}_{1}\times\mathbf{M}_{2}\right)\cdot\mathbf{N}\right]^{2}\\
 & +\left(\frac{4\lambda_{11}+8\lambda_{12}}{4}\right)N^{2}\left(M_{1}^{2}+M_{2}^{2}\right)^{2}\\
 & -\left(-\frac{4\lambda_{11}+8\lambda_{12}}{4}\right)N^{2}\left(M_{1}^{2}-M_{2}^{2}\right)^{2}
\end{split}
\end{equation}
with the coefficients: 
\begin{align}
\alpha & =4\int_{k}G_{h,k}G_{e_{1},k}^{2}G_{e_{2},k}\nonumber \\
\lambda_{ij} & =\int_{k}G_{h,k}^{2}G_{e_{1},k}G_{e_{2},k}G_{e_{i},k}G_{e_{j},k}\label{GL_diagrams}
\end{align}

\begin{figure}
\begin{centering}
\includegraphics[width=1\columnwidth]{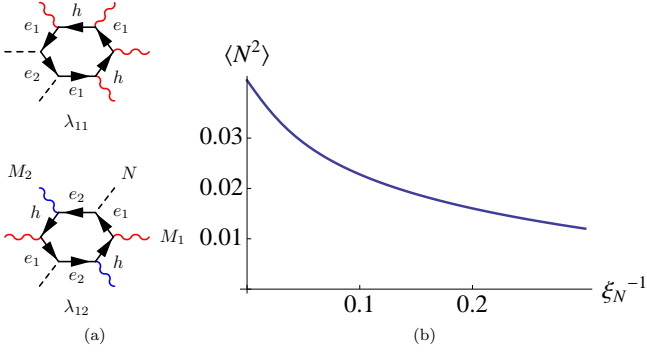} 
\par\end{centering}

\caption{(a) Feynman diagrams $\lambda_{ij}$ associated with the coupling
between the Néel collective field $\mathbf{N}$ (dashed lines) and
the magnetic order parameters $\mathbf{M}_{1}$ and $\mathbf{M}_{2}$
(wavy lines). The solid (black) lines are the Green's functions of
the corresponding bands. (b) Behavior of the Gaussian Néel fluctuations
$\left\langle N^{2}\right\rangle $ as function of the inverse Néel
correlation length $\xi_{N}^{-1}$. The Néel critical point is at
$\xi_{N}^{-1}=0$ \label{fig_Feynman_diagrams}.}
\end{figure}

The coefficients $\lambda_{ij}$ are represented diagrammatically
in Fig.~\ref{fig_Feynman_diagrams}(a). Near perfect nesting, $\alpha>0$,
reflecting the competition between the Néel and stripe states. To
study the corrections to the quartic terms of (\ref{F_general}),
denoted here by a tilde, we consider Gaussian isotropic Néel fluctuations
$\langle N_{i}N_{j}\rangle=\frac{\langle N^{2}\rangle}{3}\delta_{ij}$
and evaluate the diagrams near perfect nesting, obtaining:

\begin{align}
\frac{\tilde{u}}{u} & \approx1-0.13\,\frac{\left\langle N^{2}\right\rangle }{T_{0}^{2}}\nonumber \\
\frac{\tilde{g}}{u} & \approx0.024\left(\frac{\delta_{m}^{2}}{T_{0}^{2}}-\frac{4\left\langle N^{2}\right\rangle }{3T_{0}^{2}}\right)\nonumber \\
\frac{\tilde{w}}{u} & \approx-0.016\,\frac{\left\langle N^{2}\right\rangle }{T_{0}^{2}}\label{coeff_Neel}
\end{align}
where $T_{0}$ is the energy scale of the bare magnetic transition
temperature. Thus, when Néel fluctuations are strong enough compared
to the energy scale of the ellipticity of the electron pockets, $\left\langle N^{2}\right\rangle >\delta_{m}^{2}/2$,
the leading instability of the system is towards the non-uniform magnetic
state ($0<\tilde{g}<-\tilde{w}$), which preserves the tetragonal
symmetry of the system and induces a simultaneous checkerboard charge
order. Notice that, in the Gaussian approximation, $\left\langle N^{2}\right\rangle \propto\int_{\mathbf{q}}\chi_{N}(\mathbf{q})$
does not diverge at the Néel critical point (see Fig.~\ref{fig_Feynman_diagrams}(b)),
so this non-uniform magnetic state is not guaranteed to appear (see
also Appendix A).

The possible existence of this intermediate state between an itinerant
stripe-type state and a localized Néel phase is the main result of
this paper. We note that a similar result also holds when the Néel
instability takes place at temperatures higher than the one where
the conduction electrons order magnetically, i.e. where $N^{2}\rightarrow\left\langle N\right\rangle ^{2}$.
Note also that this approximation breaks down near the critical region
of the Néel transition, where higher-order terms may be necessary.

\section{Experimental manifestations: reconstructed electronic spectrum}

The most prominent experimental signature of the non-uniform state
is the absence of orthorhombic distortion (i.e. no splitting of the
lattice Bragg peaks) and the presence of magnetic Bragg peaks at $\mathbf{Q}_{1}=\left(\pi,0\right)$
and $\mathbf{Q}_{2}=\left(0,\pi\right)$. Indeed, this is what x-ray
and neutron diffraction experiments find in the $\mathrm{Ba(Fe_{1-x}Mn_{x})_{2}As_{2}}$
compounds for $x\gtrsim0.1$ \cite{Kim10}. However, these features
are also consistent with the non-collinear state. This is the ground
state when $g<0$ -- which may in fact be accomplished by the Néel
fluctuations, see Eq. (\ref{coeff_Neel}) -- and $w>0$, which would
require other mechanisms than Néel fluctuations \cite{Berg10}. The
key property that distinguishes between the non-uniform and non-collinear
tetragonal magnetic states is the existence of an induced checkerboard
charge order in the former, $\rho_{\mathbf{Q}_{N}}\propto\mathbf{M}_{1}\cdot\mathbf{M}_{2}$.
Because $\mathbf{Q}_{N}$ coincides with a Bragg peak of the two-Fe
unit cell, detecting this secondary order via x-ray may be challenging.
However, local probes such as STM could detect this type of charge
order. NMR could also distinguish the non-uniform and non-collinear
states, since in the former half of the sites display zero averaged
magnetization, while in the latter every site is magnetic.

We emphasize that magnetic Bragg peaks at both momenta $\mathbf{Q}_{1}=\left(\pi,0\right)$
and $\mathbf{Q}_{2}=\left(0,\pi\right)$ are also expected in the
stripe state, due to the formation of domains. This makes it difficult
to distinguish between the stripe and non-uniform states using only
neutron diffraction data. Furthermore, relying only on the absence
of orthorhombic distortion to make this distinction could be an issue
depending on the limitations imposed by the x-ray experimental resolution
-- see for instance Ref. \cite{Mn_Griffiths,Gastiasoro14}. In this
regard, absence of shear modulus softening above $T_{\mathrm{mag}}$
would provide strong evidence for a tetragonal magnetic state \cite{Rafael_prl,Yoshizawa12,Kontani11,Fernandes13_shear}.
Alternatively, the properties of the electronic spectrum in the magnetic
state could be used to differentiate between the stripe and non-uniform
states.

To obtain the reconstructed Fermi surface in the non-uniform and striped
states, we start with the five-orbital tight-binding model of Ref.
\cite{Graser09}, with the Hamiltonian: 
\begin{equation}
H_{0}=\sum_{mn}\sum_{\boldsymbol{k}\sigma}c_{m,\mathbf{k}\sigma}^{\dagger}\left(t_{mn}+\epsilon_{m}\delta_{mn}-\mu\delta_{mn}\right)c_{n,\mathbf{k}\sigma}
\end{equation}
where $\sigma$ is the spin index, and $m,n=1...5$ label the five
$d$-orbitals of the Fe atom. $t_{mn},\epsilon_{m}$ are the hopping
parameters and onsite energies given in Ref. \cite{Graser09}. The
chemical potential is $\mu=0$ for the undoped compound, corresponding
to an occupation number of $n=6$.

The presence of non-zero magnetic order parameters $\mathbf{M}_{1}$
and $\mathbf{M}_{2}$ gives rise to to the term: 
\begin{equation}
\begin{split}H_{\mathrm{mag}} & =\sum_{i=1,2}\sum_{m}\sum_{\mathbf{k}\alpha\beta}c_{m,\mathbf{k}\alpha}^{\dagger}\left(\mathbf{M}_{i}\cdot\boldsymbol{\sigma}_{\alpha\beta}\right)c_{m,\mathbf{k}+\mathbf{Q}_{i},\beta}+\\
 & \kappa\sum_{m}\sum_{\mathbf{k}\alpha}c_{m,\mathbf{k}\alpha}^{\dagger}\left(\mathbf{M}_{1}\cdot\mathbf{M}_{2}\right)c_{m,\mathbf{k}+\mathbf{Q}_{1}+\mathbf{Q}_{2},\alpha}+\mathrm{h.c.}
\end{split}
\end{equation}
where we considered only intra-orbital magnetic order parameters \cite{Plonka13},
assumed for simplicity to be of equal amplitude. $\kappa$ is a coupling
constant that determines the amplitude of the higher-order harmonic
generated when both $\mathbf{M}_{1}$ and $\mathbf{M}_{2}$ are non-zero
-- which gives rise to the checkerboard charge order. In our calculations,
we found that the reconstructed Fermi surface does not depend strongly
on the choice of $\kappa$.

The reconstructed band structure for the stripe and non-uniform orders
can be obtained by diagonalizing the full Hamiltonian $H=H_{0}+H_{\mathrm{mag}}$
adjusting the chemical potential $\mu$ under the constraint of fixed
occupation number $n=6$. To diagonalize the Hamiltonian, we introduce
the Nambu operators: 
\[
\psi_{m,\mathbf{k}\sigma}^{\dagger}=\left(\begin{array}{cccc}
c_{m,\mathbf{k}\sigma}^{\dagger} & c_{m,\mathbf{k}+\mathbf{Q}_{1}\sigma}^{\dagger} & c_{m,\mathbf{k}+\mathbf{Q}_{2}\sigma}^{\dagger} & c_{m,\mathbf{k}+\mathbf{Q}_{1}+\mathbf{Q}_{2}\sigma}^{\dagger}\end{array}\right)
\]

The order parameters couple different elements in Nambu space: $\mathbf{M}_{i}$
couples $c_{m,\mathbf{k}\sigma}^{\dagger}$ to $c_{m,\mathbf{k}+\mathbf{Q}_{i}\sigma'}$
and $c_{m,\mathbf{k}+\mathbf{Q}_{i}\sigma}^{\dagger}$ to $c_{m,\mathbf{k}+\mathbf{Q}_{1}+\mathbf{Q}_{2}\sigma'}$,
while $\mathbf{M}_{1}\cdot\mathbf{M}_{2}$ couples $c_{m,\mathbf{k}+\mathbf{Q}_{1}\sigma}^{\dagger}$
to $c_{m,\mathbf{k}+\mathbf{Q}_{2}\sigma'}$ and $c_{m,\mathbf{k}\sigma}^{\dagger}$
to $c_{m,\mathbf{k}+\mathbf{Q}_{1}+\mathbf{Q}_{2}\sigma'}$. For the
$(\pi,0)$ stripe order, $\mathbf{M}_{2}=0$ and $\mathbf{M}_{1}=M\hat{\mathbf{x}}$,
and the magnetic unit cell is given by $-\pi/2\leq k_{x}\leq\pi/2$
and $-\pi\leq k_{y}\leq\pi$. For the non-uniform magnetic order,
$\mathbf{M}_{1}=\mathbf{M}_{2}=\frac{M}{\sqrt{2}}\hat{x}$, where
the factor of $\sqrt{2}$ is introduced to keep the total order parameter
$\sqrt{M_{1}^{2}+M_{2}^{2}}$ the same as in the striped case. The
magnetic unit cell is given in this case by $-\pi/2\leq k_{x},k_{y}\leq\pi/2$.

\begin{figure}
\begin{centering}
\includegraphics[width=1\columnwidth]{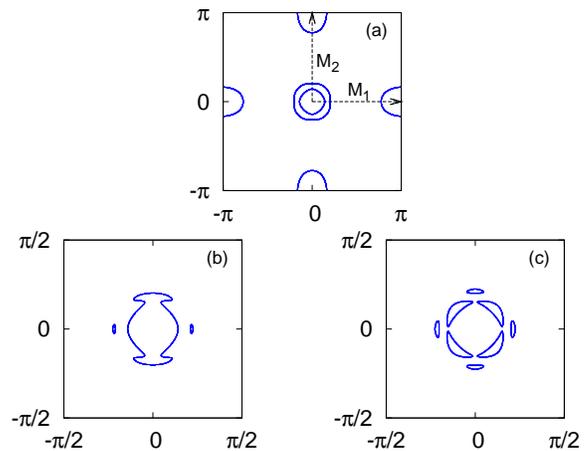} 
\par\end{centering}

\caption{Reconstructed Fermi surfaces near the center of the Brillouin zone
in the presence of $\mathbf{Q}_{1}=\left(\pi,0\right)$ stripe-type
magnetic order (b) and non-uniform magnetic order (c). The Fermi surface
in the paramagnetic state is shown in (a), with the tight-binding
parameters of Ref. \cite{Graser09}. \label{fig_reconstructed_FS} }
\end{figure}

In Fig. \ref{fig_reconstructed_FS}, we present the reconstructed
Fermi surface around the center of the magnetic Brillouin zone for
both magnetic ground states. In the paramagnetic phase, the Fermi
surface consists of two concentric hole pockets at the center of the
Brillouin zone and two elliptical pockets centered at the momenta
$\mathbf{Q}_{1}=\left(\pi,0\right)$ and $\mathbf{Q}_{2}=\left(0,\pi\right)$.
In the striped state, we find that for reasonable values of the magnetic
order parameter ($M\approx60$ meV), one of the hole pockets remains
unhybridized while the other hole pocket hybridizes with the folded
electron pocket, giving rise to ``Dirac cones'' -- the small reconstructed
pockets along the stripe modulation direction. This is in general
agreement with previous theoretical and experimental results \cite{DHLee09,Knolle11,Plonka13}.
On the other hand, for the non-uniform state, each of the two hole
pockets hybridize with one of the two electron pockets. As a result,
there remains only small reconstructed pockets \cite{Vafek13}. Unlike
the small pockets that appear in the stripe state case, four of these
pockets appear along the $\mathbf{Q}_{1}+\mathbf{Q}_{2}=\left(\pi,\pi\right)$
direction, a unique signature of the double-$\mathbf{Q}$ non-uniform
magnetic order.

\section{Coexistence between tetragonally-symmetric magnetism and superconductivity}

An intriguing observation in the $\mathrm{Ba(Fe_{1-x}Mn_{x})_{2}As_{2}}$
compounds is the absence of superconductivity, despite the fact that
the magnetic transition is suppressed down to $50$ K. In the closely
related compounds $\mathrm{Ba(Fe_{1-x}Co_{x})_{2}As_{2}}$, for instance,
one observes coexistence between superconductivity and magnetism for
similar values of $T_{\mathrm{mag}}$ \cite{FernandesPRB10}. It has
been pointed out that the Néel fluctuations in $\mathrm{Ba(Fe_{1-x}Mn_{x})_{2}As_{2}}$
effectively suppress the leading $s^{+-}$ pairing instability and
instead promote d-wave pairing \cite{Fernandes13}. Besides this effect,
the possible change in the magnetic ground state also has an impact
on the outcome of the competition between long-range magnetic order
and superconductivity.

\begin{figure}
\centering{} \includegraphics[width=0.4\textwidth]{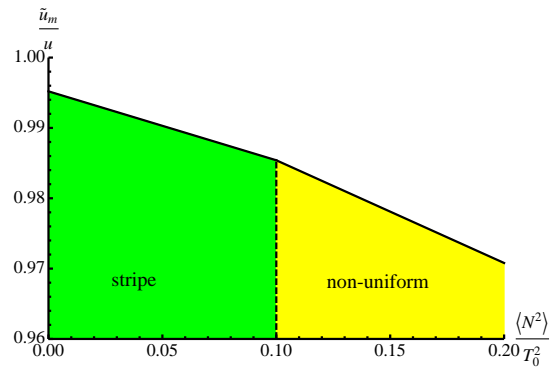}
\caption{\label{coexistence} Effective quartic magnetic coefficient $\tilde{u}_{m}$
as function of the amplitude of Néel fluctuations $\left\langle N^{2}\right\rangle $.
The condition for coexistence between superconductivity and long-range
magnetism is $\frac{\gamma}{\sqrt{u_{s}}}<\sqrt{\tilde{u}_{m}}$,
implying that the phase space for the coexistence state is reduced
as Néel fluctuations become stronger. In this plot we used $\frac{\delta_{m}^{2}}{T_{0}^{2}}=0.2$.}
\end{figure}

Within the phenomenological model (\ref{F_general}), this competition
is described by the additional Ginzburg-Landau terms:

\begin{equation}
\tilde{F}=F+F_{SC}+\frac{\gamma}{2}\Delta^{2}\left(M_{1}^{2}+M_{2}^{2}\right)\label{F_SC}
\end{equation}
where $\Delta$ is the superconducting order parameter and $\gamma>0$
is a coupling constant that can be derived directly from the microscopic
Hamiltonian $H_{0}+H_{I}$ \cite{FernandesPRB10_long,Vorontsov09}.
The superconducting free energy is given by the usual form:

\begin{equation}
F_{\mathrm{SC}}=\frac{a_{s}}{2}\Delta^{2}+\frac{u_{s}}{4}\Delta^{4}\label{aux_F_SC}
\end{equation}
$ $with $a_{s}\propto T-T_{c}$ and $u_{s}>0$. To determine whether
long-range magnetic order and superconductivity can coexist, we minimize
the free energy (\ref{F_SC}) with respect to $\Delta$ and check
whether the renormalized quartic coefficient of $M$ is positive.
In general, coexistence takes place when $\frac{\gamma}{\sqrt{u_{s}}}<\sqrt{\tilde{u}_{m}}$,
where the effective parameter $\tilde{u}_{m}$ is given by $\tilde{u}_{m}=\tilde{u}-\tilde{g}$
for the striped state and $\tilde{u}_{m}=\tilde{u}-\left|\tilde{w}\right|$
for the non-uniform state. Using our results from Eq. (\ref{coeff_Neel}),
we plot in Fig.~\ref{coexistence} the value of this effective parameter
$\tilde{u}_{m}$ as a function of the amplitude of the Néel fluctuations
for both stripe and non-uniform magnetic states. As shown in the figure,
$\tilde{u}_{m}$ decreases as Néel fluctuations become stronger, restricting
the phase space for which coexistence between superconductivity and
long-range magnetism is achieved, $\frac{\gamma}{\sqrt{u_{s}}}<\sqrt{\tilde{u}_{m}}$.
Therefore, stronger Néel fluctuations make it difficult for a coexistence
state with either stripe or non-uniform states to be realized.

\section{Concluding remarks}

In summary, we have shown that an unusual non-uniform tetragonal magnetic
state consisting of a coherent combination of $\mathbf{Q}_{1}=\left(\pi,0\right)$
and $\mathbf{Q}_{2}=\left(0,\pi\right)$ orders can be realized in
the iron pnictides as a result of the interplay between itinerant
magnetism promoted by the nesting features of the Fermi surface and
Néel-type fluctuations promoted by local moments. This non-uniform
state induces a checkerboard charge order and a reconstruction of
the electronic spectrum, both of which can be detected experimentally.
We argue that our findings may explain the experimental observation
of a tetragonal magnetic state displaying Bragg peaks at $\mathbf{Q}_{1}=\left(\pi,0\right)$
and $\mathbf{Q}_{2}=\left(0,\pi\right)$ in doped $\mathrm{Ba(Fe_{1-x}Mn_{x})_{2}As_{2}}$,
as well as the absence of coexisting superconductivity in these compounds.
Besides $\mathrm{Ba(Fe_{1-x}Mn_{x})_{2}As_{2}}$, a tetragonal magnetic
state has also been reported in $\mathrm{(Ba_{1-x}Na)Fe_{2}As_{2}}$
\cite{Osborn13}, and possibly in 122 compounds under pressure \cite{Taillefer12,Jin13},
but whether Néel fluctuations are also present in these systems remains
to be seen. The existence of such tetragonal magnetic states also
imposes important constraints on the mechanism of magnetism in the
iron pnictides, as they imply that tetragonal symmetry breaking is
not a necessary condition to achieve long-range magnetic order.

We thank A. Chubukov, I. Eremin, A. Goldman, J. Knolle, A. Kreyssig,
R. McQueeney, A. Millis, J. Schmalian, and G. Tucker for fruitful
discussions.

\appendix

\section{Gaussian Néel fluctuations}

Here we show how $\left\langle N^{2}\right\rangle $ is obtained within
a Gaussian approximation. The action for the Néel field $\mathbf{N}$
can be written as: 
\begin{equation}
S_{\text{Néel}}\left[\mathbf{N}\right]=\frac{1}{2}\int_{q}\chi_{N,q}^{-1}N^{2}+\int_{x}\frac{u}{4}N^{4}\label{Ap1}
\end{equation}
where $q=(\boldsymbol{q},\nu_{n})$ denotes both momentum and bosonic
Matsubara frequency $\nu_{n}=2\pi nT$, and $x=(\boldsymbol{r},\tau)$.
For a classical transition in a strongly anisotropic system, the Néel
susceptibility takes the form 
\begin{equation}
\chi_{N,q}^{-1}=r_{0}+q_{\parallel}^{2}+\eta_{z}\sin^{2}q_{z}\label{Ap2}
\end{equation}
where $r_{0}\propto T-T_{\text{Néel}}$ and $\eta_{z}$ is the inter-plane
coupling. Following Ref. \cite{Fernandes12}, the quartic term can
be decoupled by an auxiliary field $\psi$: 
\begin{equation}
S_{\text{eff}}[\mathbf{N},\psi]=\frac{1}{2}\int_{q}\chi_{q}^{-1}N^{2}-\int_{x}\frac{1}{4u}\psi^{2}+\frac{1}{2}\int_{x}\psi N^{2}\label{Ap3}
\end{equation}

Minimization with respect to $\psi$ gives $\langle N^{2}\rangle=\langle\psi\rangle/u$.
In the absence of long-range Néel order, the $\mathbf{N}$ field can
be directly integrated out, yielding the effective action: 
\begin{equation}
S_{\text{eff}}=-\frac{\psi^{2}}{4u}+\frac{3}{2}\int_{q}\text{ln}(\chi_{q}^{-1}+\psi)\label{Ap4}
\end{equation}

Minimization with respect to $\psi$ gives: 
\begin{equation}
\psi=3u\int_{q}\frac{1}{\chi_{q}^{-1}+\psi}\label{Ap5}
\end{equation}

Explicit evaluation then yields: 
\begin{equation}
\psi=\bar{u}\text{ ln }\frac{2\Lambda}{\sqrt{r_{0}+\psi}+\sqrt{r_{0}+\psi+\eta_{z}}}\label{Ap6}
\end{equation}
 where $\bar{u}=3uT/(2\pi)$, and $\Lambda$ is the upper cutoff.
In Figure 2b of the paper, the parameters used were $\bar{u}/\Lambda^{2}=0.01$
and $\eta_{z}/\Lambda^{2}=0.001$. The correlation length is given
by $\xi_{N}=\left(r_{0}+\psi\right)^{-1/2}$ and diverges at the Néel
transition.


\begin{thebibliography}{10}
\bibitem{Scalapino_RMP} D. J. Scalapino, Rev. Mod. Phys. \textbf{84},
1383 (2012).

\bibitem{reviews_pairing} P. J. Hirschfeld, M. M. Korshunov, and
I. I. Mazin, Rep. Prog. Phys. \textbf{74}, 124508 (2011); A. V. Chubukov,
Annu. Rev. Cond. Mat. Phys. \textbf{3}, 57 (2012).

\bibitem{LaMnPO_Kotliar} J. W. Simonson, Z. P. Yin, M. Pezzoli, J.
Guo, J. Liu, K. Post, A. Efimenko, N. Hollmann, Z. Hu, H.-J. Lin,
C. T. Chen, C. Marques, V. Leyva, G. Smith, J. W. Lynn, L. Sun, G.
Kotliar, D. N. Basov, L. H. Tjeng, and M. C. Aronson, PNAS \textbf{109},
1815 (2012).

\bibitem{Kim11} M. G. Kim, R. M. Fernandes, A. Kreyssig, J. W. Kim,
A. Thaler, S. L. Bud'ko, P. C. Canfield, R. J. McQueeney, J. Schmalian,
and A. I. Goldman, Phys. Rev. B \textbf{83}, 134522 (2011).

\bibitem{Birgeneau11} C. R. Rotundu and R. J. Birgeneau, Phys. Rev.
B \textbf{84}, 092501 (2011).

\bibitem{reviews} K. Ishida, Y. Nakai and H. Hosono, J. Phys. Soc.
Japan \textbf{78}, 062001 (2009); D. C. Johnston, Adv. Phys. \textbf{59},
803 (2010); J. Paglione and R. L. Greene, Nature Phys. \textbf{6},
645 (2010); P. C. Canfield and S. L. Bud'ko, Annu. Rev. Cond. Mat.
Phys. \textbf{1}, 27 (2010); H. H. Wen and S. Li, Annu. Rev. Cond.
Mat. Phys. \textbf{2}, 121 (2011).

\bibitem{Uchida10} M. Nakajima, S. Ishida, K. Kihou, Y. Tomioka,
T. Ito, Y. Yoshida, C. H. Lee, H. Kito, A. Iyo, H. Eisaki, K. M. Kojima,
and S. Uchida, Phys. Rev. B \textbf{81}, 104528 (2010).

\bibitem{Liu10} C. Liu, T. Kondo, R. M. Fernandes, A. D. Palczewski,
E. D. Mun, N. Ni, A. N. Thaler, A. Bostwick, E. Rotenberg, J. Schmalian,
S. L. Bud'ko, P. C. Canfield, and A. Kaminski, Nature Phys. \textbf{6},
419 (2010).

\bibitem{Andersen11} O. K. Andersen and L. Boeri, Annalen der Physik
\textbf{1}, 8 (2011).

\bibitem{Mn_synthesis} Y. Singh, A. Ellern, and D. C. Johnston, Phys.
Rev. B \textbf{79}, 094519 (2009).

\bibitem{Mn_electronic_structure} J. An, A. S. Sefat, D. J. Singh,
and M. H. Du, Phys. Rev. B \textbf{79}, 075120 (2009).

\bibitem{Mn_K_doped_ARPES} A. Pandey, R. S. Dhaka, J. Lamsal, Y.
Lee, V. K. Anand, A. Kreyssig, T. W. Heitmann, R. J. McQueeney, A.
I. Goldman, B. N. Harmon, A. Kaminski, and D. C. Johnston, Phys. Rev.
Lett. \textbf{108}, 087005 (2012).

\bibitem{Mn_Heisenberg_model} D. C. Johnston, R. J. McQueeney, B.
Lake, A. Honecker, M. E. Zhitomirsky, R. Nath, Y. Furukawa, V. P.
Antropov, and Y. Singh, Phys. Rev. B \textbf{84}, 094445 (2011).

\bibitem{Mn_NMR} Y. Texier, Y. Laplace, P. Mendels, J. T. Park, G.
Friemel, D. L. Sun, D. S. Inosov, C. T. Lin, J. Bobroff, EPL \textbf{99,}
17002 (2012).

\bibitem{Mn_Mossbauer} X. Ma, J. Bai, Z. Li, J. Wan, H. Pang, and
F. Li, J. Phys.: Condens. Matter \textbf{25}, 135703 (2013).

\bibitem{Mn_doped_1111} R. Frankovsky, H. Luetkens, F. Tambornino,
A. Marchuk, G. Pascua, A. Amato, H.-H. Klauss, and D. Johrendt, Phys.
Rev. B \textbf{87}, 174515 (2013).

\bibitem{Mn_localized_photoemission} H. Suzuki, T. Yoshida, S. Ideta,
G. Shibata, K. Ishigami, T. Kadono, A. Fujimori, M. Hashimoto, D.
H. Lu, Z.-X. Shen, K. Ono, E. Sakai, H. Kumigashira, M. Matsuo, and
T. Sasagawa, Phys. Rev. B \textbf{88}, 100501(R) (2013).

\bibitem{Mn_correlations_Schmalian} Y. X. Yao, J. Schmalian, C. Z.
Wang, K. M. Ho, and G. Kotliar, Phys. Rev. B \textbf{84}, 245112 (2011).

\bibitem{Mn_correlations_Imada} T. Misawa, K. Nakamura, and M. Imada,
Phys. Rev. Lett. \textbf{108}, 177007 (2012).

\bibitem{Kim10} M. G. Kim, A. Kreyssig, A. Thaler, D. K. Pratt, W.
Tian, J. L. Zarestky, M. A. Green, S. L. Bud'ko, P. C. Canfield, R.
J. McQueeney, and A. I. Goldman, Phys. Rev. B \textbf{82}, 220503(R)
(2010).

\bibitem{Chandra90} P. Chandra, P. Coleman, and A. I. Larkin, Phys.
Rev. Lett. \textbf{64}, 88 (1990).

\bibitem{Lorenzana08} J. Lorenzana, G. Seibold, C. Ortix, and M.
Grilli, Phys. Rev. Lett. \textbf{101}, 186402 (2008).

\bibitem{Eremin10} I. Eremin and A. V. Chubukov, Phys. Rev. B 81,
024511 (2010).

\bibitem{Tesanovic11} J. Kang and Z. Tesanovic, Phys. Rev. B \textbf{83},
020505 (2011).

\bibitem{Brydon11} P. M. R. Brydon, J. Schmiedt, and C. Timm, Phys.
Rev. B \textbf{84}, 214510 (2011).

\bibitem{Fernandes12} R. M. Fernandes, A. V. Chubukov, J. Knolle,
I. Eremin, and J. Schmalian, Phys. Rev. B \textbf{85}, 024534 (2012).

\bibitem{Mn_neutron} G. S. Tucker, D. K. Pratt, M. G. Kim, S. Ran,
A. Thaler, G. E. Granroth, K. Marty, W. Tian, J. L. Zarestky, M. D.
Lumsden, S. L. Bud'ko, P. C. Canfield, A. Kreyssig, A. I. Goldman,
and R. J. McQueeney, Phys. Rev. B \textbf{86}, 020503(R) (2012).

\bibitem{Balatsky10} A. V. Balatsky, D. N. Basov, and J.-X. Zhu,
Phys. Rev. B \textbf{82}, 144522 (2010).

\bibitem{Graser09} S. Graser, T. A. Maier, P. J. Hirschfeld, and
D. J. Scalapino, New J. Phys. \textbf{11}, 025016 (2009).

\bibitem{Maiti10} S. Maiti and A. V. Chubukov, Phys. Rev. B \textbf{82},
214515 (2010).

\bibitem{Bascones_phase_diagram} M. J. Calderon, G. Leon, B. Valenzuela,
and E. Bascones, Phys. Rev. B \textbf{86}, 104514 (2012).

\bibitem{Dagotto_phase_diagram} Q. Luo and E. Dagotto, arXiv:1308.3426.

\bibitem{Johannes09} M. D. Johannes and I. Mazin, Phys. Rev. B \textbf{79},
220510(R) (2009).

\bibitem{Fernandes13} R. M. Fernandes and A. J. Millis, Phys. Rev.
Lett. \textbf{110}, 117004 (2013).

\bibitem{ESR} P. F. S. Rosa, T. M. Garitezi, C. Adriano, T. Grant,
Z. Fisk, R. R. Urbano, R. M. Fernandes, and P. G. Pagliuso, J. Appl.
Phys. \textbf{115}, 17D702 (2014); P. F. S. Rosa \emph{et al}, unpublished.

\bibitem{BaCu2As2_1} D. J. Singh, Phys. Rev. B \textbf{79}, 153102
(2009).

\bibitem{BaCu2As2_2} V. K. Anand, P. Kanchana Perera, Abhishek Pandey,
R. J. Goetsch, A. Kreyssig, and D. C. Johnston, Phys. Rev. B \textbf{85},
214523 (2012).

\bibitem{disorder1} M. G. Vavilov and A. V. Chubukov, Phys. Rev.
B \textbf{84}, 214521 (2011).

\bibitem{disorder2} R. M. Fernandes, M. G. Vavilov, and A. V. Chubukov,
Phys. Rev. B \textbf{85}, 140512(R) (2012). 

\bibitem{Berg10} E. Berg, S. A. Kivelson, and D. J. Scalapino, Phys.
Rev. B \textbf{81}, 172504 (2010).

\bibitem{Mn_Griffiths} D. S. Inosov, G. Friemel, J. T. Park, A. C.
Walters, Y. Texier, Y. Laplace, J. Bobroff, V. Hinkov, D. L. Sun,
Y. Liu, R. Khasanov, K. Sedlak, Ph. Bourges, Y. Sidis, A. Ivanov,
C. T. Lin, T. Keller, and B. Keimer, Phys. Rev. B \textbf{87}, 224425
(2013).

\bibitem{Gastiasoro14} M. N. Gastiasoro and B. M. Andersen, arXiv:1403.3324.

\bibitem{Rafael_prl} R. M. Fernandes, L. H. VanBebber, S. Bhattacharya,
P. Chandra, V. Keppens, D. Mandrus, M. A. McGuire, B. C. Sales, A.
S. Sefat, and J. Schmalian, Phys. Rev. Lett. \textbf{105}, 157003
(2010).

\bibitem{Yoshizawa12} M. Yoshizawa, D. Kimura, T. Chiba, A. Ismayil,
Y. Nakanishi, K. Kihou, C.-H. Lee, A. Iyo, H. Eisaki, M. Nakajima,
and S. Uchida, J. Phys. Soc. Jpn. \textbf{81}, 024604 (2012).

\bibitem{Kontani11} H. Kontani, T. Saito, and S. Onari, Phys. Rev.
B \textbf{84}, 024528 (2011).

\bibitem{Fernandes13_shear} R. M. Fernandes, A. E. Böhmer, C. Meingast,
and J. Schmalian, Phys. Rev. Lett. \textbf{111}, 137001 (2013).

\bibitem{DHLee09} Y. Ran, F. Wang, H. Zhai, A. Vishwanath, and D.-H.
Lee, Phys. Rev. B \textbf{79}, 014505 (2009).

\bibitem{Knolle11} J. Knolle, I. Eremin, and R. Moessner, Phys. Rev.
B \textbf{83}, 224503 (2011).

\bibitem{Vafek13} V. Cvetkovic and O. Vafek, arXiv:1304.3723.

\bibitem{Plonka13} N. Plonka, A. F. Kemper, S. Graser, A. P. Kampf,
and T. P. Devereaux, arXiv:1308.6248.

\bibitem{FernandesPRB10} R. M. Fernandes, D. K. Pratt, W. Tian, J.
Zarestky, A. Kreyssig, S. Nandi, M. G. Kim, A. Thaler, N. Ni, P. C.
Canfield, R. J. McQueeney, J. Schmalian, and A. I. Goldman, Phys.
Rev. B \textbf{81}, 140501(R) (2010).

\bibitem{FernandesPRB10_long} R. M. Fernandes and J. Schmalian, Phys.
Rev. B \textbf{82}, 014521 (2010).

\bibitem{Vorontsov09} A. B. Vorontsov, M. G. Vavilov, and A. V. Chubukov,
Phys. Rev. B \textbf{79}, 060508(R) (2009).

\bibitem{Osborn13} S. Avci, O. Chmaissem, S. Rosenkranz, J. M. Allred,
I. Eremin, A. V. Chubukov, D.-Y. Chung, M. G. Kanatzidis, J.-P. Castellan,
J. A. Schlueter, H. Claus, D. D. Khalyavin, P. Manuel, A. Daoud-Aladine,
and R. Osborn, arXiv:1303.2647.

\bibitem{Taillefer12} E. Hassinger, G. Gredat, F. Valade, S. Rene
de Cotret, A. Juneau-Fecteau, J.-Ph. Reid, H. Kim, M. A. Tanatar,
R. Prozorov, B. Shen, H.-H. Wen, N. Doiron-Leyraud, and L. Taillefer,
Phys. Rev. B \textbf{86}, 140502 (2012).

\bibitem{Jin13} J. J. Wu, Jung-Fu Lin, X. C. Wang, Q. Q. Liu, J.
L. Zhu, Y. M. Xiao, P. Chow, and Changqing Jin, PNAS \textbf{110},
17263 (2013). \end{thebibliography}
\end{document}